\documentclass[conference,a4paper]{IEEEtran}
\IEEEoverridecommandlockouts

\usepackage{cite}
\usepackage{amsmath,amssymb,amsfonts}
\usepackage{algorithmic}
\usepackage{graphicx}
\usepackage{textcomp}
\usepackage{xcolor}
\def\BibTeX{{\rm B\kern-.05em{\sc i\kern-.025em b}\kern-.08em
    T\kern-.1667em\lower.7ex\hbox{E}\kern-.125emX}}
\begin{document}

\title{Self-Optimizing Control of Continuous Processes Based on Reinforcement Learning\\

}

\author{%
\IEEEauthorblockN{\small Ziqi Zhuo}
\IEEEauthorblockA{\small \textit{College of Control Science}\\
\small \textit{and Engineering}\\
\small Zhejiang University\\
\small HangZhou, China\\
\small ZqZhuo@zju.edu.cn}
\hspace{-1.8em}
\and
\IEEEauthorblockN{\small Junghui Chen}
\IEEEauthorblockA{\small \textit{Department of Chemical}\\
\small \textit{Engineering}\\
\small Chung-Yuan Christian Univ.\\
\small Taoyuan, R.O.C.\\
\small jason@wavenet.cycuedu.tw}
\hspace{-1.8em}
\and
\IEEEauthorblockN{\small Lei Xie}
\IEEEauthorblockA{\small \textit{College of Control Science}\\
\small \textit{and Engineering}\\
\small Zhejiang University\\
\small HangZhou, China\\
\small lxie@zju.edu.cn}
\hspace{-1.8em}
\and
\IEEEauthorblockN{\small Hongye Su}
\IEEEauthorblockA{\small \textit{College of Control Science}\\
\small \textit{and Engineering}\\
\small Zhejiang University\\
\small HangZhou, China\\
\small hysu69@zju.edu.cn}
}

\maketitle

\begin{abstract}
This paper addresses the Self-Optimizing Control (SOC) problem in industrial continuous processes and proposes a Reinforcement-Learning (RL)-based SOC approach to improve dynamic performance under high-frequency disturbances. In the proposed framework, the SOC controlled variable structure is embedded in the Actor network, and reward functions are designed based on economic indicators. Through interaction with the environment, the RL agent optimizes controlled variables while implicitly considering implementability and steady-state uniqueness. Online fine-tuning is further introduced to alleviate model mismatch. Experiments on a continuous stirred-tank reactor with disturbances compare the proposed RL-based SOC method with the Objective-Guided Controlled Variable Learning Approach based on steady-state data. The results show that the RL method achieves improved dynamic performance under real-time disturbances, generates smooth controlled variable outputs without explicit regularization, reduces hyperparameter-tuning complexity, and enhances adaptability through online adjustment. Overall, the proposed RL-based SOC approach provides an effective solution for nonlinear process control and offers a promising reference for future studies involving multiple disturbances, multiple operating conditions, and model-free scenarios.
\end{abstract}


\section{Introduction}
In industrial control, setpoint tracking is commonly employed. However, optimizing setpoints under disturbances remains a significant challenge for both industry and academia. Real-Time Optimization (RTO) is a widely used approach to address this issue. In RTO, measured disturbances are incorporated before solving an optimization problem to obtain optimal setpoints, which are implemented by lower-level controllers through setpoint tracking\cite{chachuatAdaptationStrategiesRealtime2009}. While effective, RTO for high-frequency disturbances is challenging and may lead to control deviations or economic losses.

Self-Optimizing Control (SOC) was proposed to address the limitations of economic performance under high-frequency disturbances\cite{jaschkeSelfoptimizingControlSurvey2017}. Traditional setpoint tracking maintains controlled variables near given setpoints, typically simple and specific variables such as temperature. SOC introduces a new framework by defining controlled variables as linear or nonlinear combinations of the original individual variables and assigning them fixed setpoints. The controller only needs to regulate these designed variables to the setpoints, achieving satisfactory economic performance without repeated optimization. Compared with RTO, SOC deliberately selects an economically acceptable operating point rather than the fully optimal one, which effectively mitigates economic losses caused by high-frequency disturbances.

The evolution of SOC began with locally linear approximations, in which the controlled variable was designed as $c = \boldsymbol{Hy}$, where $\boldsymbol{y}$ denotes the vector of measured variables and $\boldsymbol{H}$ is a weighting vector used to form a weighted sum of the different components of $\boldsymbol{y}$, while the mechanistic model and economic cost function were expanded around the operating point using first- or second-order Taylor series. The objective is to minimize closed-loop economic loss by optimizing $\boldsymbol{H}$. However, the performance of this locally linear SOC degrades when the process deviates significantly from the nominal operating point. To overcome this, global SOC ($gSOC$) was developed to minimize economic loss over the entire operating space. For strongly nonlinear processes, the controlled variable was designed in nonlinear forms as $c = \Phi(\boldsymbol{y})$ , where $\Phi$ denotes a nonlinear mapping. Additionally, to address variations in the active constraint set, two nonlinear SOC design approaches, namely regression-based and optimization-based approaches, are proposed based on the existence, under certain conditions, of perfectly global self-optimizing controlled variables regardless of changing active constraints; these approaches are collectively termed $g^2SOC$, where the controlled variable is defined as $c = h(\theta, \boldsymbol{y}) = u - f(\theta, \boldsymbol{y})$\cite{yeGeneralizedGlobalSelfOptimizing2023}. Here, $h$ and $f$ are nonlinear mappings, with $f$ derived from steady-state data, and $u$ denotes the measurable manipulated variable. Despite this, $g^2SOC$ may still face challenges in closed-loop performance and computational efficiency. Recent works have introduced implementability and uniqueness regularization during training, named the Objective-Guided Controlled Variable Learning Approach (OGCVL), achieving improved performance in nonlinear processes \cite{zhouGeneralizedGlobalSelfOptimizing2025}.

For OGCVL, the training data are derived from steady-state conditions, ensuring economic performance at these steady-state points. However, real-time disturbances prevent the system from remaining at steady state. In addition, the involved weights must be adjusted to balance multiple objectives. Moreover, conventional SOC is mostly designed offline; any model mismatch or change in disturbance distribution after deployment requires re-optimization. This motivates the need for a method that considers both steady-state and off-steady-state economic performance, requires no additional regularization training for $f(\theta, \boldsymbol{y})$, and allows online real-time fine-tuning.

This paper proposes a Reinforcement-Learning (RL)-based SOC for continuous processes, treating $f(\theta, \boldsymbol{y})$ as the Actor network in a RL agent. By leveraging economic-reward-based design and learning through the Critic network, the Actor network policy is iteratively improved, automatically optimizing $f(\theta, \boldsymbol{y})$ and enabling online fine-tuning. The main contributions of this work are:

\begin{itemize}
\item \textbf{Enhancing dynamic performance of SOC}: By interacting mechanistic models, the proposed method extends SOC for continuous processes to consider not only steady-state performance but also dynamic trajectories under disturbances. Through cumulative rewards in continuous control, the dynamic performance distribution is incorporated into SOC design.
\item \textbf{Implicit regularization during nonlinear $f(\theta, \boldsymbol{y})$ training}: Previous SOC approaches required explicit regularization terms to ensure implementability and solution uniqueness, necessitating careful weight tuning. Here, iterative strategy updates within a mechanistic stepwise environment implicitly account for these regularizations, yielding improved closed-loop performance.
\item \textbf{Online adaptability of RL-based SOC}: Unlike conventional offline SOC designs, the RL-based SOC framework can collect data online and perform real-time adjustments to compensate for model mismatch, enhancing practical applicability.
\end{itemize}

\section{Problem Statement}
\subsection{Overview of Self-Optimizing Control}\label{AA}
In industrial continuous processes, optimization and control are typically structured in two layers: an upper-level RTO layer and a lower-level controller (e.g., PID). The steady-state optimization procedure can be formulated as follows:

\begin{equation}
\begin{aligned}
& \min_{u} \; J(u, d) \\
& \text{s.t.} \\
& \quad \boldsymbol{y} = F(u, d) \\
& \quad \phi(u, d) \leq 0 \\
& \quad \psi(u, d) = 0 \\
& \quad u \in \mathcal{U}, \; d \in \mathcal{D}
\end{aligned}
\end{equation}
where, $J(u,d)$ denotes the economic cost function, $\boldsymbol{y} = F(u,d)$ represents the output measurements, and $\phi(u,d)$ and $\psi(u,d)$ correspond to the inequality and equality constraints, respectively. The manipulated variable $u$ and the disturbance $d$ are bounded as $u \in \mathcal{U}$ and $d \in \mathcal{D}$. 

The RTO layer solves this optimization problem to determine the optimal manipulated variable $u^*$ and the corresponding optimal output $\boldsymbol{y}^*$, which is then passed to the lower-level controller as the setpoint for tracking. Typically, RTO requires the system to reach steady state before re-estimating disturbances and performing optimization using a steady-state model. For high-frequency disturbances, however, waiting for steady state leads to suboptimal operation and economic losses, making RTO ineffective. In such cases, SOC is employed to perform optimization under the specific disturbance. 

SOC selects a suitable controlled variable structure and its setpoint based on the economic cost function and disturbances, denoted as $c = h(\theta, \boldsymbol{y})$ and $c_s$, where $h(\cdot, \cdot)$ represents the chosen controlled variable structure, $\theta$ are structural parameters, and $\boldsymbol{y}$ are output measurements. Controlling $c$ to the setpoint $c_s$ enhances overall economic performance. Common choices for $h(\cdot, \cdot)$ include the linear structure $c = \boldsymbol{Hy}$ and the nonlinear structure $c = \mathcal{NN}_{\boldsymbol{\theta}}(\boldsymbol{y})$. Accordingly, the SOC problem can be formulated as follows:

\begin{equation}
\begin{aligned}
& \min_{\theta, \, c_s} \; J(h(\theta,\boldsymbol{y}),c_s) \\
& \text{s.t.} \\
& \quad c_s = h(\theta, \boldsymbol{y}) \\
& \quad \boldsymbol{y} = F(u, d) \\
& \quad \phi(u, d) \leq 0 \\
& \quad \psi(u, d) = 0 \\
& \quad u \in \mathcal{U}, \; d \in \mathcal{D}
\end{aligned}
\end{equation}

The nonlinear mapping considered in this work is $c = h(\theta, \boldsymbol{y}) = u - f(\theta, \boldsymbol{y})$.

\begin{figure*}[t]  
    \centering
    \includegraphics[width=0.8\textwidth]{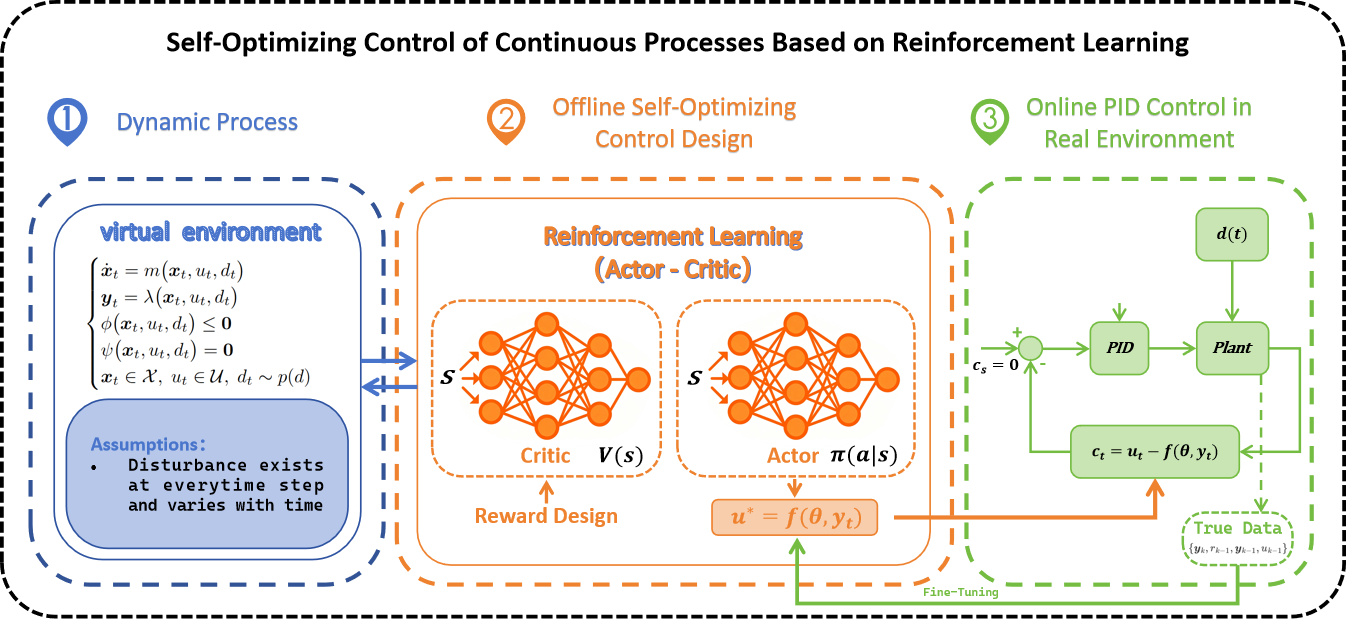}  
    \caption{Self-Optimizing Control Based on Reinforcement Learning Algorithm Framework}
    \label{Algorithm Framework.png}
\end{figure*}

\subsection{Problems to Be Solved}
Previous SOC methods for continuous processes face several challenges:
\begin{itemize}
\item \textbf{Neglect of dynamic performance in distributed processes}: Conventional SOC approaches often focus solely on steady-state optimization. In practice, disturbances render the controlled process dynamic and stochastic. Incorporating the performance of such distributed dynamic processes into SOC design is a key challenge addressed in this work.
\item \textbf{Tuning of regularization weights for nonlinear}: Previous methods introduce regularization terms to ensure implementability and uniqueness. Balancing these weights typically requires extensive trial-and-error. Efficiently achieving implementability and steady-state uniqueness constitutes another challenge addressed here.
\item \textbf{Economic losses due to model mismatch}: After offline design of controlled variable structures and parameters, discrepancies between the actual process and the simulation model can arise, necessitating retraining of $f(\theta, \boldsymbol{y})$. Implementing online fine-tuning to mitigate model mismatch and enhance practical applicability is another problem tackled in this work.
\end{itemize}

The method proposed in this paper provides concrete solutions to these challenges.

\section{Methodology}
As shown in Fig.~\ref{Algorithm Framework.png}, the algorithm framework of this work consists of two parts: offline SOC design using a mechanistic model and online fine-tuning using real data, and these two parts will be described in detail in the following sections.
\subsection{Offline SOC Design}
\subsubsection{Dynamic Process}
Offline SOC design relies on a mechanistic model, and this work models the dynamics of stochastic processes under distributed disturbances as:
\begin{equation}
\begin{cases}
\dot{\boldsymbol x}_t = m\big(\boldsymbol x_t, u_t, d_t\big) \\
\boldsymbol y_t = \lambda\big(\boldsymbol x_t, u_t, d_t\big) \\
\phi\big(\boldsymbol x_t, u_t, d_t\big) \le \boldsymbol 0 \\
\psi\big(\boldsymbol x_t,u_t,d_t\big) = \boldsymbol 0 \\
\boldsymbol x_t \in \mathcal{X},\;u_t \in \mathcal{U},\;d_t \sim p(d)
\end{cases}
\end{equation}
where $\boldsymbol x_t$ and $\dot{\boldsymbol x}_t$ denote the state and its derivative at time $t$, $u_t$ is the manipulated variable, $d_t$ represents a time-varying disturbance, and $\boldsymbol y_t$ is the measured output.

The controlled variable is structured as:
\begin{equation}
c = h(\theta, \boldsymbol{y}) = u - f(\theta, \boldsymbol{y})
\end{equation}
where $f(\theta, \boldsymbol{y})$, implemented as an Multi-Layer Perceptron (MLP), is the part to be optimized, and $u$ is the actual manipulated variable.

\subsubsection{Reinforcement Learning}

This work considers dynamic performance under disturbance distributions, and the implementability and steady-state uniqueness regularizations are incorporated implicitly. By introducing RL, part of $f(\theta, \boldsymbol{y})$ is treated as the Actor in an Actor-Critic network:
\begin{equation}
c = u - f(\theta, \boldsymbol{y}), \quad f(\theta, \boldsymbol{y}) \text{ from Actor Network}
\end{equation}

The RL algorithm used in this work is Proximal Policy Optimization (PPO). PPO is an RL algorithm that constrains single-step updates to avoid performance degradation due to large policy changes \cite{schulmanProximalPolicyOptimization2017}. Its clipped objective is:

\begin{equation}
\begin{aligned}
&L_t^{\mathrm{CLIP}}(\theta) \\
&= \mathbb{E}_t \Bigg[ \min \Big( r_t(\theta) \hat{A}_t,\quad \operatorname{clip}\big(r_t(\theta), 1-\epsilon, 1+\epsilon\big) \hat{A}_t \Big) \Bigg]
\end{aligned}
\end{equation}
where $r_t(\theta) = \frac{\pi_\theta(u_t|\boldsymbol{y}_t)}{\pi_{\theta_\mathrm{old}}(u_t|\boldsymbol{y}_t)}$.\\

The complete PPO loss in the Actor-Critic framework is:
\begin{equation}
L_t(\theta) = \mathbb{E}_t \left[ L_t^{\mathrm{CLIP}}(\theta) - c_1 L_t^{\mathrm{VF}}(\theta) + c_2 S[\pi_\theta](\boldsymbol{y}_t) \right]
\end{equation}
where $L_t^{\mathrm{VF}}(\theta)$ is the value function loss, and $S[\pi_\theta](\boldsymbol{y}_t)$ is the policy entropy term.

A reward function based on economic indicators guides exploration and exploitation in a mechanistic step environment. The RL agent optimizes parameters $\theta$ via the advantage function, thereby improving dynamic SOC performance while implicitly accounting for implementability and uniqueness. After optimization, $u^* = f(\theta, \boldsymbol{y})$ and $c_s = 0$, completing the offline SOC design.

It is worth noting that the trained optimal mapping $u^* = f(\theta, \boldsymbol{y})$ applies not only to steady-state $\boldsymbol{y}$, but also to non-steady-state $\boldsymbol{y}$. Therefore, for any $t \in [0,T]$, the optimal mapping can be expressed as $u_t^* = f(\theta, \boldsymbol{y}_t)$.

\subsection{Online Deployment and Fine-Tuning}
\subsubsection{Online Deployment}
After RL training,  $u^* = f(\theta, \boldsymbol{y})$ and $c_s = 0$, and the controlled variable structure is:
\begin{equation}
c = u - f(\theta, \boldsymbol{y})
\end{equation}
where $f(\theta, \boldsymbol{y})$ is the Actor network of the PPO agent. 

Online, a PI controller maintains $c$ near the setpoint, yielding SOC that is robust to disturbances and exhibits superior dynamic performance.

\subsubsection{Online Fine-Tuning}

Unfortunately, discrepancies may exist between the real process and the mechanistic model, causing $f(\theta,\boldsymbol{y})$ learned offline to be suboptimal in practice. 
To address model mismatch, real process data are collected as:
\begin{equation}
\{\boldsymbol{y}_k,r_{k-1},\boldsymbol{y}_{k-1},u_{k-1}\}
\end{equation}

These data are further utilized in the RL policy optimization algorithm for online fine-tuning, thereby enabling rapid adaptation to scenarios with model mismatch.

\section{Case Study}
\subsection{Continuous Stirred Tank Reactor}
\begin{figure}[htbp]
    \centering
    \includegraphics[width=0.75\linewidth]{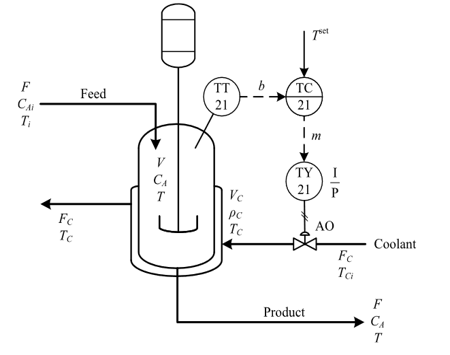}
    \caption{CSTR}
    \label{cstr}
\end{figure}

In this study, a Continuous Stirred Tank Reactor (CSTR) with disturbances is considered as the controlled process. A schematic diagram is shown in Fig.~\ref{cstr}. The dynamic equations can be found in Ref\cite{bequetteProcessControlModeling2024}.

Consider a disturbance applied to the system:
\begin{equation}
T_i(t) \sim \mathcal{N}(\mu_{T_i}=66.0, \ \sigma_{T_i}^2=8)
\end{equation}
where $T_i$ denotes the feed temperature.

Parameters in the dynamic equations are listed in Table ~\ref{tab:parameters}.

\begin{table}[htbp]  
\centering
\caption{Simulation Parameters}
\scriptsize
\setlength{\tabcolsep}{3pt}

\begin{tabular}{|c|c|c|c|}
\hline
\textbf{Parameter} & \textbf{Value} & \textbf{Unit} & \textbf{Description} \\
\hline
$dt$ & 10 & s & Simulation time step \\
$V$ & 7.08 & m$^3$ & Reactor volume \\
$V_C$ & 1.82 & m$^3$ & Jacket volume \\
$F$ & 0.0075 & m$^3$/s & Feed rate \\
$F_{C\text{max}}$ & 0.02 & m$^3$/s & Maximum flow \\
$\alpha$ & 50 & - & Valve rangeability parameter \\
$A$ & 5.40 & m$^2$ & Heat-transfer area \\
$U$ & 3550 & $\mathrm{J/(s \cdot m^2 \cdot C)}$ & Overall heat-transfer coefficient \\
$\Delta H_R$ & $-9.86\times10^7$ & J/kgmole & Heat of reaction \\
$\rho$ & 19.2 & kgmole/m$^3$ & Reactor contents density \\
$C_p$ & 181500 & J/(kgmole$\cdot$C) & Heat capacity \\
$\rho_C$ & 1000 & kg/m$^3$ & Coolant density \\
$C_{pC}$ & 4184 & J/(kg$\cdot$C) & Coolant heat capacity \\
$E$ & $1.182\times10^7$ & J/kgmole & Activation energy \\
$k_0$ & 0.0744 & $\mathrm{m^3/(s \cdot kgmole)}$ & Frequency parameter \\
$R$ & 8314.39 & J/(kgmole$\cdot$K) & Gas constant \\
\hline
\end{tabular}

\label{tab:parameters}
\end{table}

Since concentration is difficult to measure in real time, the output measurements of the CSTR are:
\begin{equation}
\boldsymbol{y} = \{
T,
T_C\}
\end{equation}
where $T$ denotes the reactor temperature, and $T_c$ denotes the jacket temperature.

The manipulated variable $u$ represents the controller output (valve opening) and ranges from 0 to 1.

\subsection{Past Method: OGCVL \& RTO}

\subsubsection{OGCVL as Baseline}
First, steady-state optimization is performed under different $T_i$ using the economic cost function:
\begin{equation}
r
=
- w_u u^2
- w_c \left(C_A - C_{A,\mathrm{ref}}\right)^2
\end{equation}
where $w_u=1.0,~w_c=1000000.0$.

The concentration deviation penalty is $10^6$ times larger than the valve opening penalty because the concentration deviation is approximately $10^{-3}$ (squared $\sim 10^{-6}$), whereas the valve opening is around 0.4 (squared $\sim 10^{-1}$). Such weighting, differing by a factor of $10^6$, is therefore appropriate.

The optimal steady-state dataset $\{T_i^k,u^k_{opt},\boldsymbol{y}^k_{opt}\}$ is then obtained. Next, $f(\theta,\boldsymbol{y})$ in $c=u-f(\theta,\boldsymbol{y})$ is modeled as an MLP with ReLU activation, and the loss function is defined as:
\begin{equation}
L_{total}=\alpha L_w+\lambda L_a+\mu L_u
\end{equation}
where:

\begin{itemize}
\item $L_w$ is the weighted loss, given by $L_w = \left\| \mathcal{L}_{uu}^{1/2} \left( u_{opt}-f(\boldsymbol{y}_{opt}, \theta) \right) \right\|_2^2$, where $\mathcal{L}_{uu}$ denotes the second derivative of the Lagrangian with respect to $u$
\item $L_a$ is the implementability regularization term, given by $L_a = \left\| f_y \mathcal{G}_u \right\|_2^2$, where $f_y$ denotes the first derivative of the nonlinear mapping $f$ with respect to $\boldsymbol{y}$, and $\mathcal{G}_u$ denotes the first derivative of the output measurement $\boldsymbol{y}$ with respect to $u$
\item $L_u$ is the solution uniqueness regularization term, given by $L_u=\sum_{j=1}^{n_y} f_{k,y}(j,:) \frac{\partial^2 y_j}{\partial u^2}$
\end{itemize}

\begin{figure}[htbp]   
    \centering
    \includegraphics[width=0.8\linewidth]{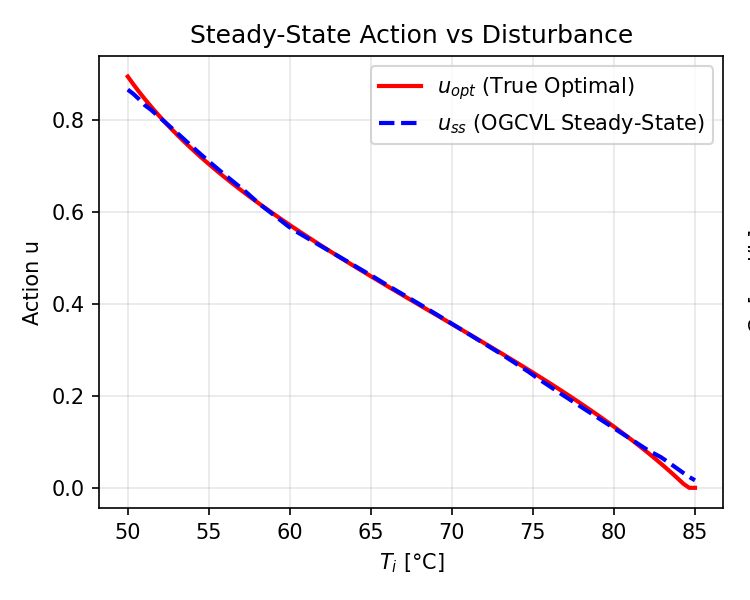}  
    \caption{Steady-state performance of the Baseline}
    \label{fig:baseline}  
\end{figure}

\begin{figure*}[htbp]
    \centering
    \setcounter{figure}{4}
    \includegraphics[width=\textwidth]{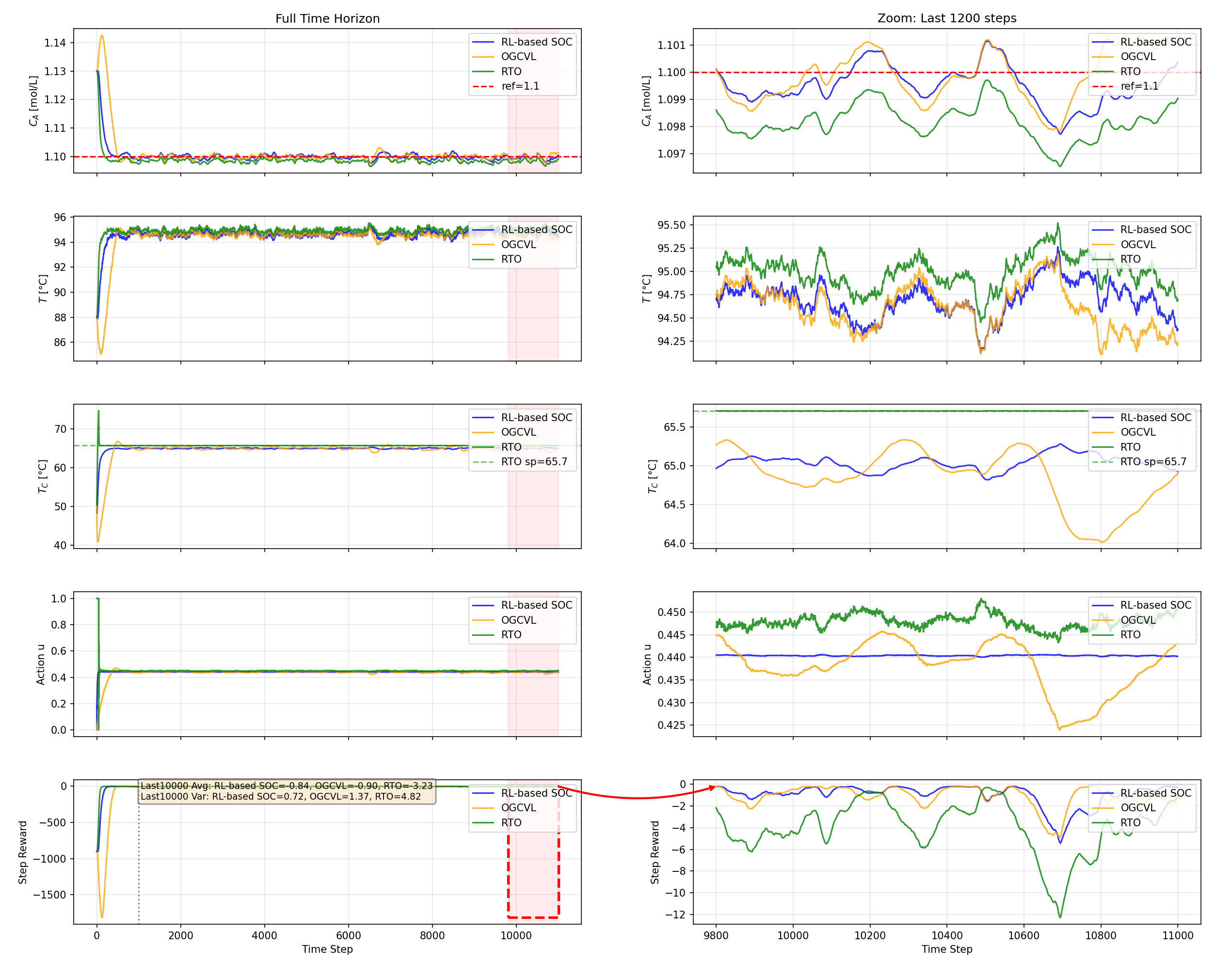}
    \caption{RL-based SOC vs Baseline vs RTO under the same $T_i$ sequence}
    \label{compare.png}
\end{figure*}

The parameters are set as $\alpha = 1.20,\ \lambda = 0.01,\ \mu = 10.0$, with normalized neural-network inputs and outputs. The trained $f(\theta,\boldsymbol{y})$ is used to construct the controlled variable, and the PI setpoint is specified as $c_s=0$.

To assess convergence of Baseline training and steady-state optimality under disturbances, $T_i$ is assumed constant long enough for the PI controller to reach steady state. Fig.~\ref{fig:baseline} compares the manipulated variable with the disturbance value: the red solid line denotes the optimal \(u\) from steady-state optimization, whereas the blue dashed line denotes the steady-state \(u\) achieved by OGCVL. The results indicate convergence close to $u_{opt}$ for nearly all $T_i$ values.

\subsubsection{RTO}
For comparison, $T_C$ was optimized using the steady-state model at $T_i=65.4$℃, yielding a setpoint of ${T_C}_{s}=65.7$ ℃ which was used to evaluate the economic performance of the RTO scheme.

\begin{figure}[htbp]
    \centering
    \setcounter{figure}{3}
    \includegraphics[width=0.8\linewidth]{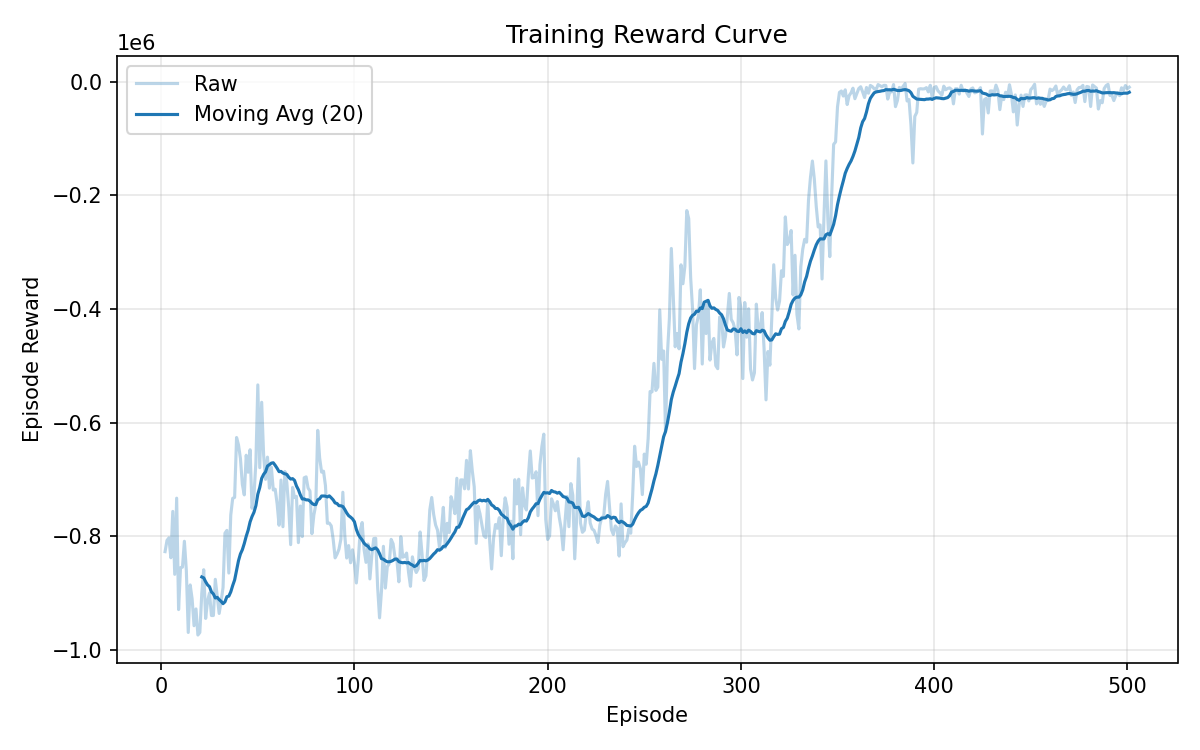}
    \caption{Training Reward Curve in RL (Offline)}
    \label{Reward}
\end{figure}

\subsection{Our Method: RL-based SOC}
\subsubsection{Offline SOC Design}
The RL agent's state space is defined as the CSTR output measurements $\boldsymbol{y}$, and the action space corresponds to the valve opening $u \in [0,1]$. The policy is defined as:
\begin{equation}
u \sim \pi_\theta(u | T, T_C)
\end{equation}

The reward function is designed as:
\begin{equation}
r = \left( - w_u u^2 - w_c \left(C_A - C_{A,\mathrm{ref}}\right)^2 \right) \cdot \frac{\exp\left(\alpha \frac{t}{T}\right)-1}{\exp(\alpha)-1}
\end{equation}
where $C_A$ denotes the concentration of the reactant in the reactor, $C_{A,\mathrm{ref}}$ is its reference value, and $\frac{\exp\left(\alpha \frac{t}{T}\right)-1}{\exp(\alpha)-1}$ is a time-dependent weighting factor that gradually increases the reward contribution, emphasizing the later near-steady-state stage rather than the initial transient stage.

The reward parameters are set as \(w_u = 1.0\), \(w_c = 10^6\), and \(\alpha = 5\), consistent with the settings used in Baseline training.

After RL training, the reward progression is shown in Fig.~\ref{Reward}.

The model achieving the highest reward is saved as $f(\theta, \boldsymbol{y})$ and used as part of $c = u - f(\theta, \boldsymbol{y})$ to implement SOC.

\subsubsection{Online Deployment}

Fig.~\ref{compare.png} compares the dynamic performance of RL-based SOC, the Baseline, and RTO under high-frequency disturbances. For SOC, $f(\theta,\boldsymbol{y})$ learned by RL or the Baseline method is embedded in $c=u-f(\theta,\boldsymbol{y})$, with a PI controller tracking $c_s=0$; RTO instead tracks the optimized steady-state setpoint of $T_C$. The figure shows time profiles of \(C_A\), \(T\), \(T_C\), \(u\), and the single-step reward, with global and zoomed views in the left and right panels, respectively. 

Performance is evaluated using the mean and variance of rewards over the last 10,000 steps. The mean rewards for RL-based SOC, OGCVL, and RTO are $-0.84$, $-0.90$, and $-3.23$, respectively, with corresponding variances of $0.72$, $1.37$, and $4.82$, indicating that RL-based SOC provides superior average performance and robustness.

\begin{figure}[htbp]
    \centering
    \setcounter{figure}{5}
    \includegraphics[width=0.7\linewidth]{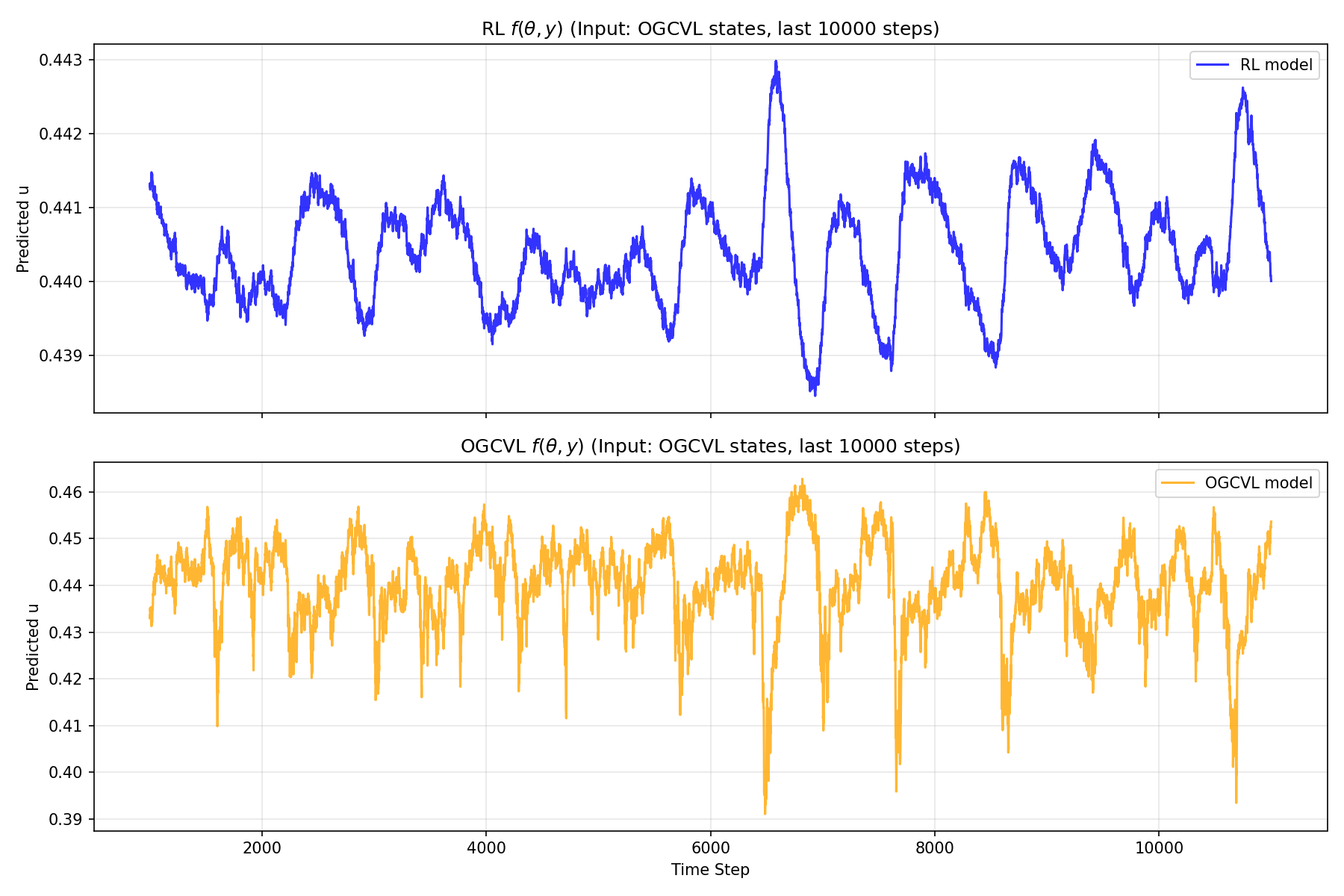}
    \caption{$f(\theta, \boldsymbol{y})$: RL-based SOC vs Baseline}
    \label{compare_model_predictions}
\end{figure}

RL-based SOC also exhibited smoother control than the Baseline. Fig.~\ref{compare_model_predictions} compares fluctuations in the output of \(f(\theta,\boldsymbol{y})\) under the same input $\boldsymbol{y}$, taken from an OGCVL segment in Fig.~\ref{compare.png}. The variation magnitude is approximately $0.005$ for RL-based SOC, compared with $0.05$ for the Baseline.

Overall, both RL-based SOC and the Baseline outperform RTO in economic performance. Notably, RL learns $f(\theta,\boldsymbol{y})$ without explicit regularization while achieving higher average rewards, lower reward variance, and smoother control than the Baseline.

\subsection{Online Fine-Tuning}

The RL-based SOC was originally trained using the mechanistic model with $V=7.08$, where satisfactory performance was achieved. To evaluate model--process mismatch, this pretrained policy is fine-tuned in the actual process environment with $V=6.90$ and compared with direct online training without pretraining. The results are shown in Fig.~\ref{continuous_500k_trajectory}.

These results indicate that RL-based SOC can be safely fine-tuned online under model--process mismatch, improving economic performance and robustness relative to direct training.

\begin{figure}[htbp]
    \centering
    \setcounter{figure}{6}
    \includegraphics[width=1.0\linewidth]{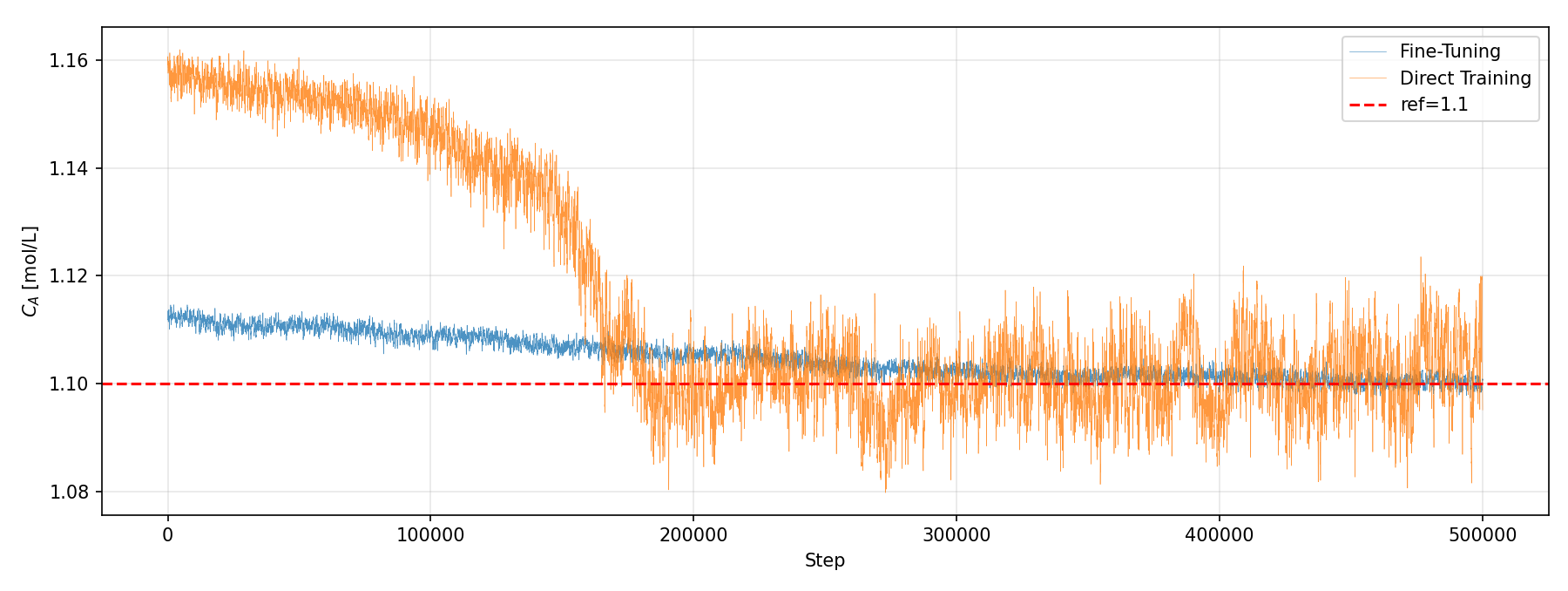}
    \vspace{-2em}
    \caption{Online Fine-Tuning vs Direct Training}
    \label{continuous_500k_trajectory}
    \vspace{-1em}
\end{figure}

\section{Conclusion}
In summary, RL-based SOC showed promising performance on the disturbed CSTR, demonstrating that RL can effectively train the nonlinear mapping $f(\theta,\boldsymbol{y})$ and improve dynamic performance under distributed disturbances. The method does not rely on highly sensitive hyperparameters and, under model mismatch, can mitigate economic losses through online fine-tuning.

Nevertheless, several limitations remain:

\begin{itemize} 
\item How to implement RL-based SOC when no mechanistic model is available, and only historical data are accessible. 
\item How to extend RL-based SOC to handle multiple disturbances and multiple operating conditions. \end{itemize}

Overall, RL-based SOC offers a feasible framework for nonlinear controlled-variable design and optimization under uncertainty in process industries. Future work should further improve its generalizability and practical applicability.

\bibliographystyle{ieeetr} 
\bibliography{reference}   

\end{document}